\documentclass[aps,pre,twocolumn,superscriptaddress]{revtex4}
\usepackage{amssymb}
\usepackage{epsfig}
\usepackage{amsmath}
\usepackage{times}
\usepackage{booktabs}
\usepackage{array}
\usepackage{color}

\begin{document}
	
\title{Social contagion under hybrid interactions}

\author{Xincheng Shu}
\affiliation{Institute of Cyberspace Security, Zhejiang University of Technology,  Hangzhou 310023, China}
\affiliation{Binjiang Institute of Artificial Intelligence, ZJUT, Hangzhou 310056, China}

\author{Man Yang}
\affiliation{Institute of Cyberspace Security, Zhejiang University of Technology,  Hangzhou 310023, China}
\affiliation{Binjiang Institute of Artificial Intelligence, ZJUT, Hangzhou 310056, China}

\author{Zhongyuan Ruan}
\email{zyuan.ruan@gmail.com}
\affiliation{Institute of Cyberspace Security, Zhejiang University of Technology,  Hangzhou 310023, China}
\affiliation{Binjiang Institute of Artificial Intelligence, ZJUT, Hangzhou 310056, China}

\author{Qi Xuan}
\affiliation{Institute of Cyberspace Security, Zhejiang University of Technology,  Hangzhou 310023, China}
\affiliation{Binjiang Institute of Artificial Intelligence, ZJUT, Hangzhou 310056, China}

\begin{abstract}
Threshold-driven models and game theory are two fundamental paradigms for describing human interactions in social systems. However, in mimicking social contagion processes, models that simultaneously incorporate these two mechanisms have been largely overlooked. Here, we study a general model that integrates hybrid interaction forms by assuming that a part of nodes in a network are driven by the threshold mechanism, while the remaining nodes exhibit imitation behavior governed by their rationality (under the game-theoretic framework). Our results reveal that the spreading dynamics are determined by the payoff of adoption. For positive payoffs, increasing the density of highly rational nodes can promote the adoption process, accompanied by a double phase transition. The degree of rationality can regulate the spreading speed, with less rational imitators slowing down the spread. We further find that the results are opposite for negative payoffs of adoption. This model may provide valuable insights into understanding the complex dynamics of social contagion phenomena in real-world social networks.
\end{abstract}

\maketitle

\section{Introduction} 
Social systems consist of agents with diverse characteristics, whose interactions can lead to notable social contagion phenomena, such as the global adoption of new technologies, products, or ideas \cite{Granovetter:1983,Karsai:2014,Lehmann:2018,Ruan:2015}. However, the nature of the interactions among agents that drives these collective behaviors remains unclear. Various mechanisms have been proposed to characterize such contagion processes, including generalized disease spreading models \cite{Anderson:1992,Satorras:2015,Moreno:2004,Dietz:1967,Barrat:2008} and threshold models \cite{Watts:2002,Gleeson:2007,Galstyan:2007,Ruan:2022,Ruan:2021}. In the first class of models, the spreading dynamics are usually driven by binary interactions, where each interaction between paired nodes is independent. In contrast, threshold models account for the effects of peer pressure, assuming that a node's state is influenced by the states of multiple neighbors. Specifically, in the linear threshold model, a node changes its state only if the fraction of neighbors who have switched their state exceeds a given threshold value \cite{Watts:2002}. It is generally difficult to distinguish which mechanism is at work in real spreading scenarios. Recently, some efforts have been made to address this issue \cite{Cencetti:2023,Andres:2024}. 

On the other hand, in economics, social agents are typically assumed to be rational. Game theory provides a mathematical framework to describe the interactions among rational individuals and has been widely applied to various areas, such as ecology and epidemic spread \cite{Smith:1973,Ye:2021,Andrea:2022,Bauch:2003,Kabir:2020,Huang:2024,Szabo:2007}. Early game-theoretic studies assume that individuals are homogeneously mixed and fully rational, making decisions to achieve the most advantageous outcomes for themselves \cite{Neumann:1953}. These assumptions have been extended by considering more realistic factors, for example, the heterogeneity in the number of contacts and the tendency to imitate their contacts who appear to have adopted successful strategies \cite{Mbah:2012,Bauch:2005,Traulsen:2010,Fu:2011}.

Peer pressure effects and game-based imitation behavior may coexist in social interaction processes. However, a social contagion model that simultaneously incorporates the above two mechanisms remains lacking, despite some attempts to integrate simple and complex contagion into a unified model \cite{Min:2018,Czaplicka:2016}. In this paper, we aim to fill this gap by proposing a general model that considers this hybrid interaction form. Specifically, we assume that part of the nodes in a system tend to imitate their neighbors' behaviors based on the payoff of a decision and the rationality of individuals, while the remaining nodes follow the threshold-driven mechanism, with their dynamics determined by the average contact number and the average threshold of an individual \cite{Watts:2002}. We find that, different from the previous studies, the dynamics is determined by the payoff of adoption. Moreover, the rationality of individuals also plays a key role, particularly in the temporal evolution of the adoption process.  

This paper is organized as follows. In Section II, we introduce our model, which incorporates two distinct types of nodes, each governed by a unique interaction mechanism. In Section III, we present simulation results for various scenarios, including both positive and negative adoption payoffs. In Section IV, we provide a theoretical analysis of our model. Finally, we conclude in Section V.

\section{Model}
\begin{figure}
\epsfig{figure=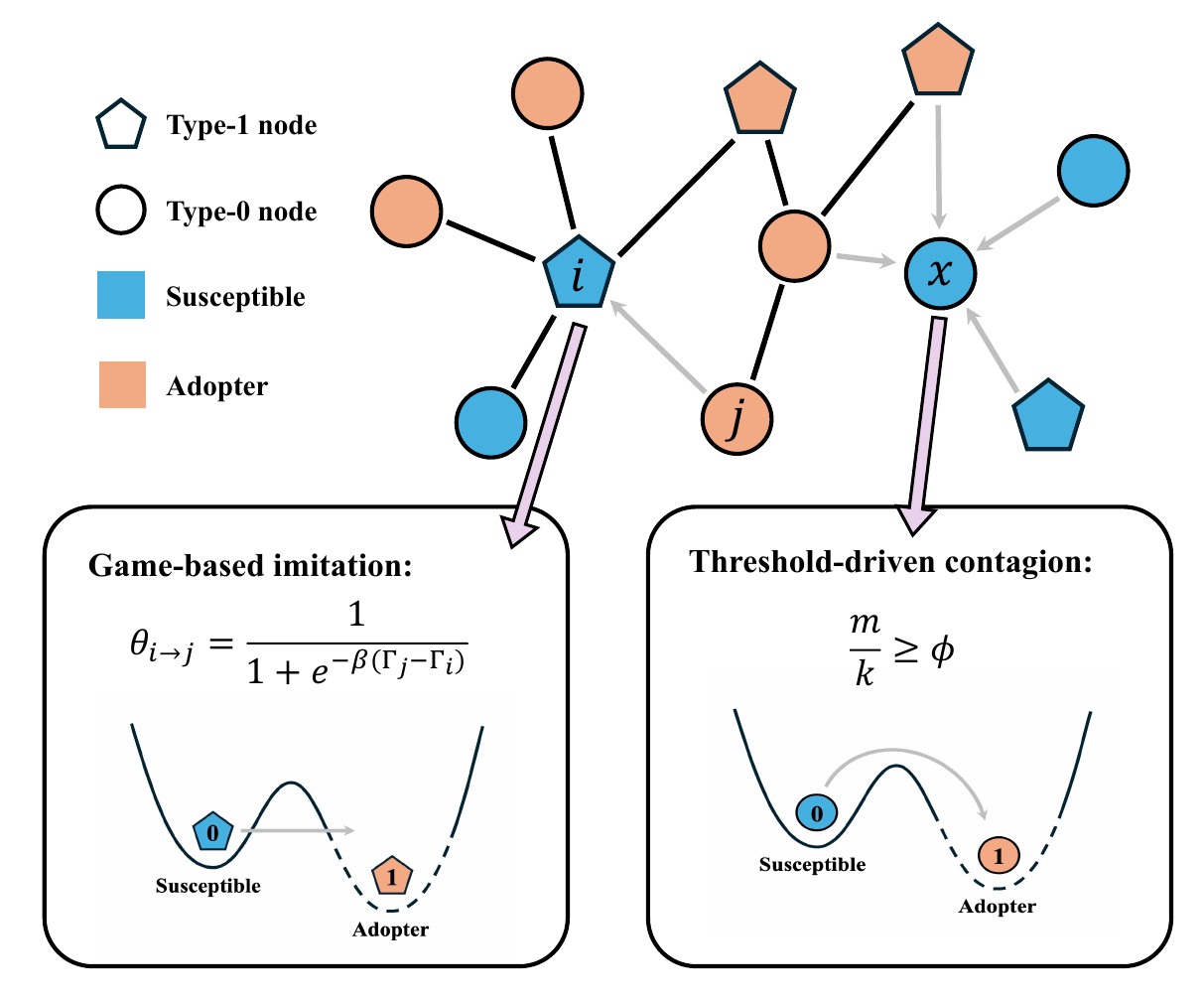,width=1.0\linewidth} \caption{(color online).  Schematic illustration of the model. There are two types of nodes in the network, each following its own specific adoption rule. Type-0 nodes (represented as pentagons) adhere to the threshold-driven mechanism, changing state only if the fraction of their adopter neighbors exceeds a given threshold value, i.e., $m/k \ge \phi$. Type-1 nodes tend to imitate others' strategies and will change their state with a certain probability, without any threshold barriers.} \label{Fig:1}
\end{figure}

We consider a graph $G(N,E)$ of $N$ nodes and $E$ edges, with an average degree $z=2E/N$. Here we mainly focus on the random ER networks \cite{Albert:2002}. Nodes in the network can be in one of the two states: susceptible or adopter. Initially, all nodes are in a susceptible state except for one randomly selected node, which is in adopter state. The system evolves as susceptible nodes transition into the adopter state according to predefined rules. Once a node becomes an adopter, it remains in that state permanently. This dynamic process could be loosely compared to a particle jumping in a double-well potential, where one well has infinitely negative energy, as illustrated in Fig. \ref{Fig:1}.

We assume there are two types of nodes obeying two different rules to switch their state. The first type of nodes (referred to as type-0 nodes, constituting a fraction of $1-r$) are sensitive to peer pressure, with their adoption behavior determined by the states of all neighbors. Specifically, each of these nodes changes its state from $0$ to $1$ if the fraction of its adopter neighbors exceeds a given threshold value $\phi$, i.e., $m_i/k_i \ge \phi $, where $k_i$ is the degree of node $i$, and $m_i$ is the number of adopters among its neighborhood. The adoption rate for a type-0 node with degree $k_i$, therefore, is
\begin{eqnarray}\label{eq:adoption-rate1}
F_{k_i,0}=  
\begin{cases}  
1, & \text{if } m_i \geq k_i\phi \\  
0, & \text{otherwise} 
\end{cases}.
\end{eqnarray}

The remaining fraction $r$ of nodes act as imitators (referred to as type-1 nodes), making adoption decisions by following the choices of others, guided by principles of game theory. We assign each node $i$ a payoff $\Gamma_i$. For susceptible nodes, $\Gamma_i$ is set to $0$, while for adopters, the payoff is defined by the difference between the benefit $b_i$ of adoption and the cost $c_i$ this node has to pay, i.e.,
\begin{eqnarray}\label{eq:payoff}
\Gamma_i=b_i-c_i.
\end{eqnarray}
For convenience, we constrain the values of $b_i$ and $c_i$ within the range of $[0, 1]$ so that they are commensurable. In reality, the benefit $b_i$ (as well as the cost $c_i$) may be contingent on various factors and vary among individuals. Here, we consider a simplified scenario in which all adopters share a uniform value, i.e., $b_i=b$ and $c_i=c$. More general considerations, such as assuming that $b_i$ and $c_i$ follow specific distributions, can be readily accommodated.

At each time step, each susceptible imitator $i$ (with $\Gamma_i=0$) selects one of its neighbors $j$ at random, and imitates its strategy (whether to adopt or not) with a probability given by the Fermi function
\begin{eqnarray}\label{eq:fermi}
\theta_{i\to j} =\frac{1}{1+e^{-\beta (\Gamma_j-\Gamma_i)}},
\end{eqnarray}
here we assume $\beta\in (-\infty,+\infty)$, which accounts for the ``rationality" of the nodes. When $\beta\to +\infty$, node $i$ is completely ``rational", signifying that it will unequivocally adopt (or reject) $j$'s strategy whenever $\Gamma_j>\Gamma_i$ (or $\Gamma_j<\Gamma_i$); conversely, as $\beta \to -\infty$, node $i$ becomes completely ``anti-rational", meaning it will do the opposite: adopt $j$'s strategy when $\Gamma_j<\Gamma_i$ and reject it when $\Gamma_j>\Gamma_i$. Under these assumptions, the adoption rate for an imitator $i$ can be explicitly calculated, i.e.,
\begin{eqnarray}\label{eq:adoption-rate2} 
F_{k_i,1}=\theta \frac{m_i}{k_i},
\end{eqnarray}
where $m_i/k_i$ denotes the probability of choosing an adopter neighbor. In particular, when $\theta=1$, the updating dynamics is equivalent to the Voter model with the recovery rate equaling to $0$ \cite{Gleeson:2013}. 

\section{Results} 
\subsection{Case of positive payoffs}
\begin{figure}
\epsfig{figure=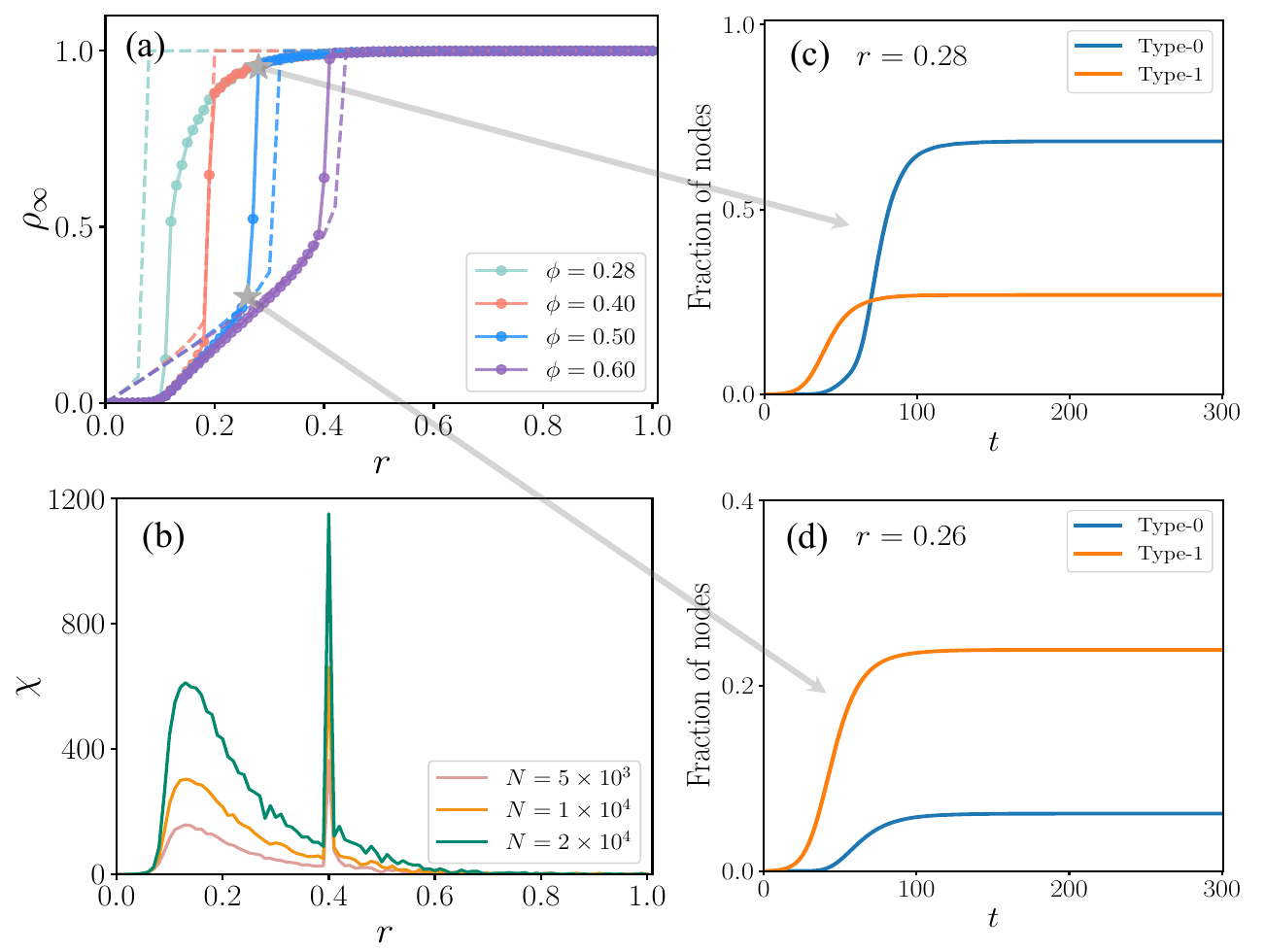,width=1.0\linewidth} \caption{(a) Final adoption density $\rho_\infty$ as a function of the fraction of imitators $r$ for different values of $\phi$. The dashed lines correspond to the theoretical results. (b) Susceptibility $\chi$ as a function of $r$ for $\phi=0.6$. (c) and (d) Time evolution of the fraction of the two types of nodes for $r=0.28$ and $0.26$, respectively, under $\phi=0.5$. The positions of the two selected values of $r$ are marked as stars on the blue line in (a). The other parameters in the simulations are chosen as $N=5000$, $z=12$, $b=1$, $c=0$, and $\beta=1000$. Results are averaged over $10^4$ realizations.} \label{Fig:2}
\end{figure}

We first consider the case $b>c$, where the benefit of adoption exceeds the cost incurred. In this context, if individuals are highly rational, i.e., $\beta \to +\infty$, the Fermi function tends towards $1$. According to Eq. (\ref{eq:adoption-rate2}), we note that the imitators will eventually become adopters as long as the number of their adopting neighbors is nonzero. Once becoming adopter, they may further contribute to the adoption of type-0 nodes, helping them to overcome the threshold barrier. This promotion effect is clearly demonstrated in Fig. \ref{Fig:2}a and c. In Fig. \ref{Fig:2}c, we illustrate how the fraction of the two types of nodes evolve over time for $r=0.28$, a condition under which global cascades can occur (i.e., the final adoption density $\rho_\infty$ is macroscopic). We find that, at early times, the imitation process dominates. After $t\approx 50$, some nodes begin to be stimulated by the threshold-driven mechanism. Eventually, the number of threshold-driven nodes surpasses and dominates the process. 

Fig. \ref{Fig:2}a shows how the final adoption density $\rho_\infty$ changes with the fraction of imitators $r$. We observe that as $r$ increases, $\rho_\infty$ reaches to a high level via two different patterns for various values of $\phi$. When $\phi$ is relatively small (for example, $\phi=0.28$ in the figure), as $r$ gradually increases, there appears one tipping point $r_c$, beyond which $\rho_\infty$ will rise from $0$ to a finite value. Here, we focus on the paired parameters $(z,\phi)$ outside the cascading window of the original Watts model, ensuring that no global cascades are possible for $r=0$. In this scenario, type-0 nodes are easily swayed to switch their state (for small $\phi$ nodes are easy to fulfill the threshold condition). Hence, a small fraction of imitators (given that they are exposed to adopter neighbors) can effectively trigger global cascades.

However, for large values of $\phi$, there appear two tipping points, suggesting a double phase transition \cite{Min:2018,Simon:2014}. To confirm this, we introduce a modified susceptibility defined as \cite{Souza:2019,Ferreira:2012,Arruda:2023}
\begin{eqnarray}\label{eq:susceptibility} 
\chi=N\frac{\langle \rho_\infty^2 \rangle-\langle \rho_\infty \rangle^2}{\langle \rho_\infty \rangle}.
\end{eqnarray}
A peak in the susceptibility $\chi$ signals the presence of a phase transition, and its position provides an estimate of the threshold. Taking $\phi=0.6$ as an example, we observe two distinct peaks in the $\chi(r)$ curve, with these peaks becoming more pronounced as the network size $N$ increases (see Fig. \ref{Fig:2}b). These findings confirm that the double phase transition is genuine.

To understand this phenomenon, notice that for large $\phi$, the type-0 nodes need a sufficient number of adopter neighbors to change state, potentially requiring more imitators to provide ``external forces" (i.e., increasing $m/k$) for them to cross the barrier. As the number of imitators in a network progressively increases, there will be a critical point $r_c^1$ above which a giant connected cluster of imitators emerges (a traditional percolation process). For random ER networks, $r_c^1=1/z$. This giant cluster is vulnerable to external disturbances: once one node in this cluster becomes adopter, it will induce all the remaining nodes in it to change state through the imitation process, whose transient time is determined by the ``rationality" of nodes $\beta$ [see Eq. (\ref{eq:fermi})]. It should be emphasized that this cluster is different from the ``vulnerable" cluster defined by Watts in \cite{Watts:2002}, which is composed of nodes with degree $k\le 1/\phi$, and its transition from susceptible to adopter state (all nodes in the cluster become adopter) is almost instantaneous. Further increasing $r$, the size of the giant cluster of imitators will continuously increase, contributing to the linear growth of the final adoption density $\rho_\infty$. In this stage (before arriving the second tipping point), the imitation process dominates, as indicated by few type-0 nodes transitioning from susceptible to adopter state (see Fig. \ref{Fig:2}d). 

\begin{figure}
\epsfig{figure=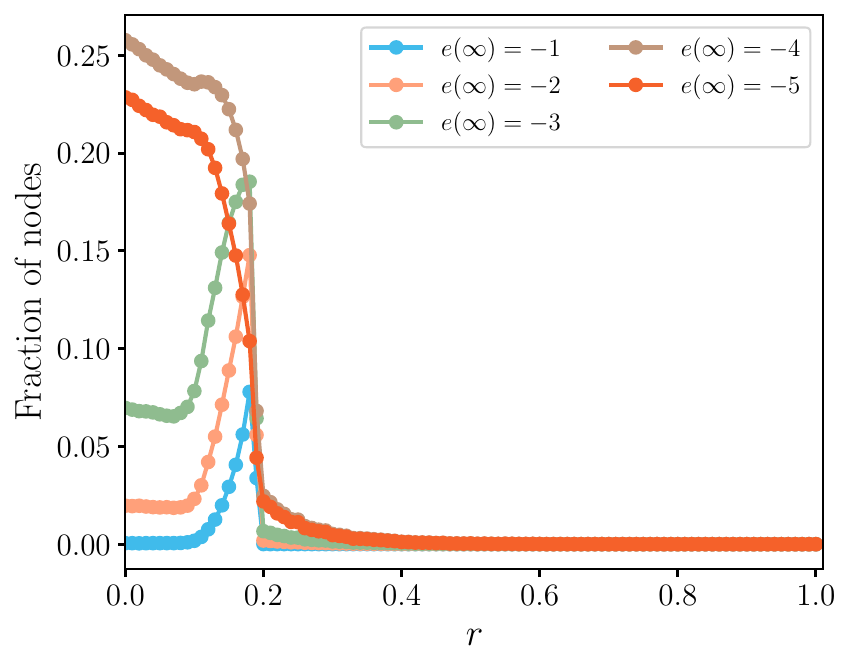,width=1.0\linewidth} \caption{Final fraction of nodes with different values of $e$ as a function of $r$, where $|e(\infty)|=k-m(\infty)$ denotes the minimal number of adopter neighbors required for a node with degree $k$ and $m$ adopting neighbors to change state. The parameters in the simulations are $N=5000$, $z=12$, $b=1$, $c=0$, $\phi=0.4$, and $\beta=1000$. Results are averaged over $10^4$ realizations.} \label{Fig:3}
\end{figure}

On the other hand, with the continuous growth of imitators, type-0 nodes become increasingly vulnerable. To quantify this, we introduce a quantity $e_i(t)=m_i(t)-k_i\phi$ for each node $i$ at time $t$, analogy to the ``energy state" of this node (where $e_i$ is an integer), and $|e_i(t)|$ denotes the ``energy" required for the node to transition from susceptible to adopter. It should be noted that nodes with $e_i(t)=-1$ are the most vulnerable, as one additional adopter neighbor is sufficient to trigger their state transition. Consequently, nodes with $e_i(t)=-2$ denotes the second most vulnerable, and so on. Fig. \ref{Fig:3} shows the fraction of nodes with different values of $e_i$ at the steady state as a function of $r$. We observe that as $r$ increases, the number of nodes in unstable states (e.g., $e_i = -1, -2$ or $-3$) grows almost synchronously. This continues until $r\approx r_c^2$, where the numbers reach their respective peaks, setting the stage for an abrupt global cascade.

\begin{figure}
\epsfig{figure=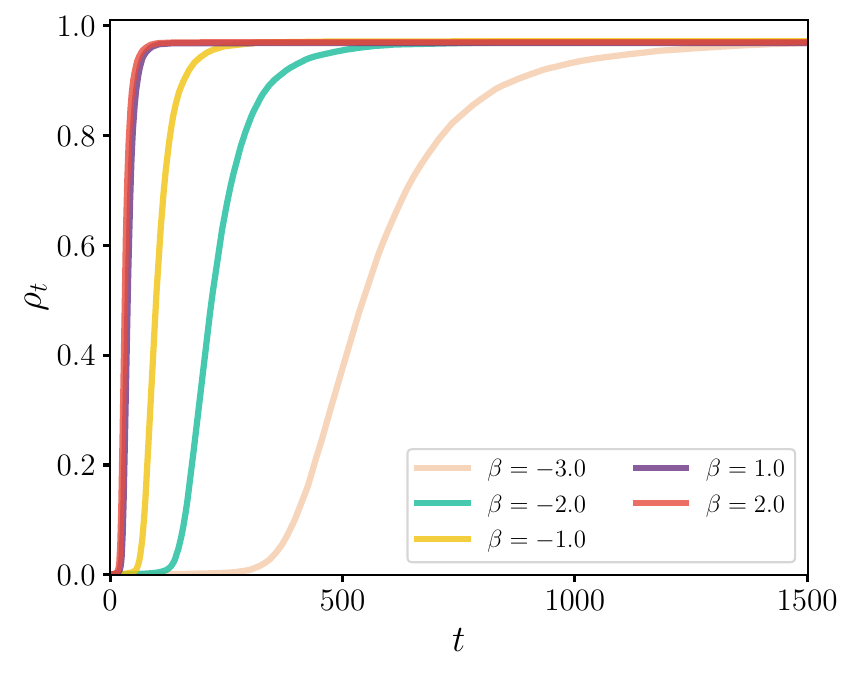,width=1.0\linewidth} \caption{Time evolution of the adoption density $\rho_t$ for varying $\beta$. Curves correspond to $N=5000$, $z=12$, $b=1$, $c=0$, $\phi=0.2$, $r=0.3$, and are averaged over $10^4$ realizations.} \label{Fig:4}
\end{figure}

Next, we explore the case where individuals are not completely rational, i.e., $\beta < +\infty$. Note that the role of $\beta$ in the dynamics is to regulate the imitation probability $\theta$ [see Eq. (\ref{eq:fermi})], which determines how quickly an imitator turns into an adopter. Therefore, it is expected that the asymptotic state of the system is unaffected by $\beta$, while the evolution process largely depends on it. Fig. \ref{Fig:4} shows how the adoption density changes with time $t$. We see that as $\beta$ increases from negative (small) values to positive (large) values, the evolution process speeds up: $\rho_t$ will grow faster to the steady state. This finding offers an alternative explanation for why, in some instances, real-world cascades may evolve slowly over time \cite{Lehmann:2018}, rather than as predicted by the Watts model, where global cascades emerge almost instantaneously.

\begin{figure}
\epsfig{figure=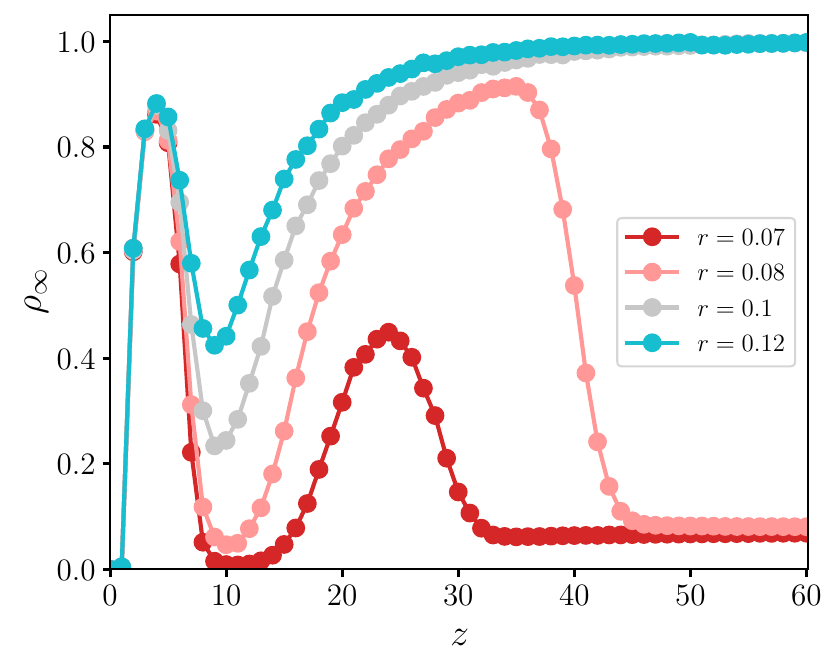,width=1.0\linewidth} \caption{Final adoption density $\rho_\infty$ as a function of the average degree $z$ for different values of $r$. Simulations correspond to $N=5000$, $b=1$, $c=0$, $\phi=0.2$, and are averaged over $10^4$ realizations.} \label{Fig:5}
\end{figure}

Finally, we investigate the impact of the average degree of the network on the spreading dynamics. Without imitators (i.e., $r=0$), as $z$ increases (moving far away from $1$), global cascades become impossible since nodes are difficult to meet the threshold condition. Therefore, we observe that the final adoption density $\rho_\infty$ decreases for large $z$. However, when imitators are introduced, the increase in link density may have an opposite effect: the likelihood of imitators being infected also increases, further promoting the spreading process. These two opposing forces may cause $\rho_\infty$ to rebound after it drops as $z$ grows, as shown in Fig. \ref{Fig:5}. It should be noted that for small values of $z$, the spreading dynamics are restricted by the network structure. In particular, for $z<1$, the network is fragmented. In this case, increasing the number of imitators has no effect on the spreading process. Consequently, we see there is a common tipping point $z=1$ for different values of $r$. 

\subsection{Case of negative payoffs}
In this section, we focus on the case $b<c$, where the adoption payoff is negative. This situation frequently occurs in real life. For example, consider the adoption of an expensive but unnecessary technology or product. Another example could be investing in a financial scheme with high risk and low expected returns. In this scenario, if individuals are highly rational, i.e., $\beta \to +\infty$, the imitators never adopt (since $\theta \to 0$), behaving as blocked nodes as studied in \cite{Ruan:2015}. These nodes may shrink the cascade window (in which global cascades can occur) and slow down the spreading process, playing an inhibitory role in the cascading dynamics. 

\begin{figure}
\epsfig{figure=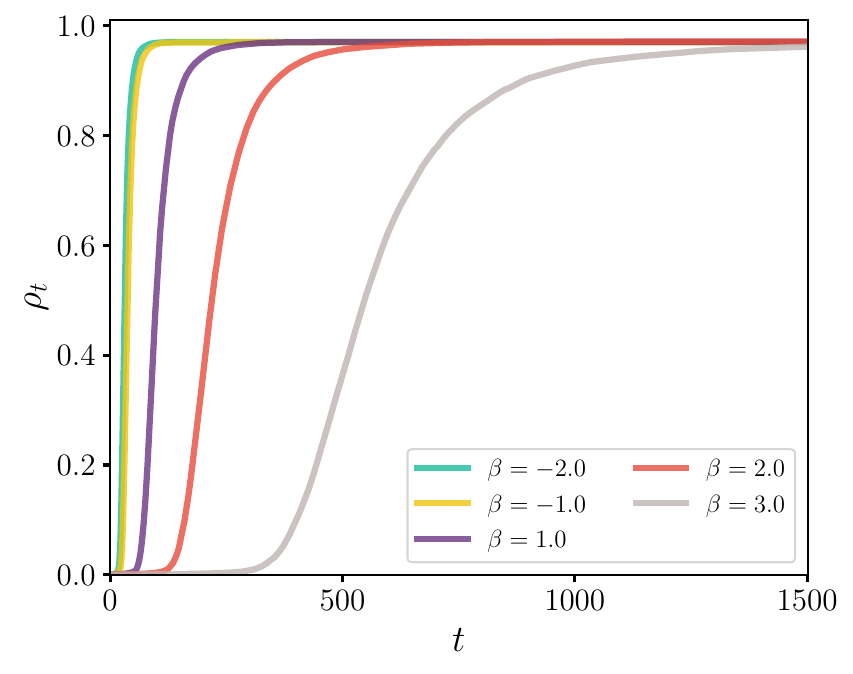,width=1.0\linewidth} \caption{Time evolution of the adoption density $\rho_t$ for varying $\beta$ in the case of $b<c$. Curves correspond to $N=5000$, $z=12$, $b=0$, $c=1$, $\phi=0.2$, $r=0.3$, and are averaged over $10^4$ realizations.} \label{Fig:6}
\end{figure}

Nevertheless, for $\beta< +\infty$, the imitators become less strictly rational and have the possibility to adopt, which may facilitate the spread. Fig. \ref{Fig:6} demonstrates the adoption density $\rho_t$ as a function of time $t$ for various $\beta$ values. The result is opposite to the case of $b>c$: for smaller $\beta$, $\rho_t$ grows more rapidly to a steady state, whereas for larger $\beta$, the evolution dynamics significantly decelerate. This finding reveals that reduced rationality can accelerate the spread of irrational behaviors or information, mirroring how absurd actions occasionally proliferate in the real world.

\section{Theoretical Analysis}
Our model can be studied analytically by extending the framework of approximate master equations (AMEs) developed by Gleeson \cite{Ruan:2015,Gleeson:2013}, particularly for $\theta=1$. For each node in the network, we introduce a vector $\mathbf{k}=(k,c)$ to describe its property, where $k=0,1,2,...k_M$ is the degree of the node, and $c=0,1$ denotes its type. In our model, $c=0$ corresponds to nodes that adopt with a threshold $\phi$ (following the threshold mechanism), while $c=1$ corresponds to nodes that prefer to imitate others' decisions. Under the mean-field assumption, nodes with the same values of $\mathbf{k}$ are considered to be equivalent. We let $p_{\mathbf{k}}$ be the probability that a randomly selected node has property vector $\mathbf{k}$. Given that the imitators (a fraction of $r$) are selected randomly in the network, we have $p_{(k,0)}=(1-r)p_k$ and $p_{(k,1)}=rp_k$, where $p_k$ is the degree distribution of the underlying network.

Let $s_{\mathbf{k},m}(t)$ be the fraction of $\mathbf{k}$-class nodes that are susceptible at time $t$ and have $m$ adopting neighbors. Correspondingly, the fraction of nodes with property vector $\mathbf{k}$ that are adopters at time $t$ is $\rho_{\mathbf{k}}=1-\sum_{m=0}^k s_{\mathbf{k},m}(t)$. The total fraction of adopters in the network at time $t$ thus is given by $\rho(t)=\sum_{\mathbf{k}} p_{\mathbf{k}}\rho_{\mathbf{k}}$, where the sum over vector $\mathbf{k}$ means $\sum_{\mathbf{k}}=\sum_k\sum_c$. Following the AMEs developed by Gleeson, the rate equation for $s_{\mathbf{k},m}(t)$ can be written as 
\begin{eqnarray}\label{eq:rate-eq}  
\frac{ds_{\mathbf{k},m}}{dt}&=&-F_{\mathbf{k},m}s_{\mathbf{k},m}-\beta_s(k-m)s_{\mathbf{k},m} \nonumber\\
&&+\beta_s(k-m+1)s_{\mathbf{k},m-1},
\end{eqnarray}
where $\beta_s(t)$ is the rate at which S-S edges transform to S-I edges, $m=0,1,...,k$, and $s_{\mathbf{k},-1}\equiv 0$.  $\beta_s$ can be calculated as the ratio between the number of edges that switch from S-S to S-I during time interval $dt$ and the number of S-S edges in the network at time $t$, 
\begin{eqnarray}\label{eq:beta_s}  
\beta_s(t)=\frac{\sum_{\mathbf{k}} p_{\mathbf{k}} \sum_m (k-m)F_{\mathbf{k},m}s_{\mathbf{k},m}}{\sum_{\mathbf{k}} p_{\mathbf{k}} \sum_m (k-m) s_{\mathbf{k},m}}.
\end{eqnarray}
Assuming that a fraction $\rho(0)$ of nodes is randomly selected as adopters at the beginning, the initial conditions for Eq. (\ref{eq:rate-eq}) are $s_{\mathbf{k},m}(0)=[1-\rho(0)]B_{k,m}[\rho(0)]$, where $B_{k,m}[\rho(0)]$ is the binomial factor
\begin{eqnarray}\label{eq:binomial}  
B_{k,m}(\rho)=\binom{k}{m} \rho^m (1-\rho)^{k-m}.
\end{eqnarray}
If the initial seed size is infinitesimally small, i.e., $\rho(0) \to 0$, we have $s_{\mathbf{k},m}(0)=B_{k,m}(0)$.

For a general network with the largest degree denoted as $k_{M}$, the number of differential equations in the system (\ref{eq:rate-eq}) is $(k_M+1)(k_M+2)$, which grows quadratically with the maximum degree. Nevertheless, the AMEs for our model can be reduced to a lower-dimensional system using the method developed by Gleeson.

To proceed, we consider three system-wide quantities: $\rho_0(t)=1-\sum_k p_k\sum_m s_{(k,0),m}(t)$, the fraction of type-0 (i.e., $c=0$) nodes that are adopters; $\rho_1(t)=1-\sum_k p_k\sum_m s_{(k,1),m}(t)$, the fraction of type-1 (i.e., $c=1$) nodes that are adopters; and $v(t)=\sum_\mathbf{k} p_{\mathbf{k}} \sum_m ms_{\mathbf{k},m}(t)/\sum_m ks_{\mathbf{k},m}(t)$, the probability that a randomly selected neighbor of a susceptible node is in adopter state. 

We then propose an ansatz solution for Eq. (\ref{eq:rate-eq}) in terms of the following forms, 
\begin{eqnarray}  
s_{(k,0),m}(t)&=&B_{k,m}[v(t)] ~~~~ \text{for}~m<k\phi, \label{eq:ansatz}\\
s_{(k,1),m}(t)&=&[1-\rho_1(t)]B_{k,m}[v(t)], \label{eq:ansatz1}
\end{eqnarray}
where $B_{k,m}(v)$ follows Eq. (\ref{eq:binomial}). It is worth noticing that in Eq. (\ref{eq:ansatz1}), we have made a rather crude mean-field assumption that the probability of a randomly chosen node in class $\mathbf{k}=(k,1)$ being susceptible is $1-\rho_1(t)$. This assumption is applicable when considering homogeneous networks. 

Using the ansatz solution, Eq. (\ref{eq:rate-eq}) can finally be reduced to three ordinary differential equations (see SM)
\begin{eqnarray}\label{eq:reduced}  
\dot{\rho}_0&=&h(v)-\rho_0, \\
\dot{\rho}_1&=&(1-\rho_1)v, \\
\dot{v}&=&g(v,\rho_1)-v, 
\end{eqnarray}
where
\begin{eqnarray}\label{eq:g}  
g(v,\rho_1)&=&(1-r)\sum_k \frac{kp_k}{z}\sum_{m\ge k\phi}B_{k-1,m}(v) \nonumber\\
&&+r[1-(1-\rho_1)(1-v+\frac{v}{z})],
\end{eqnarray}
and
\begin{eqnarray}\label{eq:h_v}  
h(v)=\sum_k p_k\sum_{m\ge k\phi}B_{k,m}(v).
\end{eqnarray}

The above equations can be solved numerically to obtain $\rho_0(t)$, $\rho_1(t)$ and $v(t)$. The fraction of adopters at time $t$ can be calculated as
\begin{eqnarray}\label{eq:rho}  
\rho&=&1-\sum_{\mathbf{k}} p_{\mathbf{k}}\sum_m s_{\mathbf{k},m}(t) \nonumber\\
&=& 1- (1-r)\sum_k p_k\sum_m s_{(k,0),m}-r\sum_kp_k \sum_m s_{(k,1),m} \nonumber\\
&=& (1-r)\rho_0+r\rho_1.
\end{eqnarray}

In Fig. \ref{Fig:2}, we show the theoretical results of how $\rho_\infty$ varies with $r$ for different values of $\phi$ (see the dashed lines therein). We observe a reasonable agreement with the simulation results. However, it should be noted that the theory fails to capture the first tipping point observed in the simulations. This discrepancy arises from the mean-field approximation of the irreversible binary-state dynamic process on random networks. The differential equations predict that the adoption density will cumulatively increase over time for any average degree $z$. However, in simulations, when $z<1$, contagion becomes impossible due to the fragmentation of the underlying network.

\section{Discussion} 
Motivated by the observation that agents in real social networks exhibit diverse characteristics, we have proposed a simple networked contagion model that incorporates two kinds of nodes, each following its own state transition rule. Specifically, we assume that some nodes are sensitive to peer pressure and switch their state according to the threshold condition, while the remaining nodes behave as imitators, following the game-based formalism. We have introduced a minimalist game framework, assuming that (i) all adopters have a uniform payoff, and (ii) susceptible imitators follow the strategy of one of their neighbors at random according to the Fermi function. Our results show that the spreading dynamics depend strongly on the adoption payoffs. For positive payoffs, rational imitators are more likely to follow the decisions of the adopters. In this scenario, increasing the fraction of imitators in the system can effectively facilitate the cascading process, potentially leading to double phase transitions. The speed of spreading is regulated by the rationality of the imitators: the more rational the nodes, the faster the growth of adopters. However, for negative payoffs, the result is the opposite. Highly rational nodes will play an inhibitory role in the cascade dynamics. Furthermore, as rationality decreases, the adoption density speeds up. 

It is important to emphasize that incorporating the game-based formalism into the complex social contagion process offers several distinct advantages. First, it effectively captures realistic factors like individual rationality and adoption payoffs, which are typically crucial in real-world decision-making processes. Second, the model is highly flexible, accommodating different types of individuals—such as blocked nodes—simply by adjusting its parameters. Finally, it enables control over the evolution of the process without affecting the system's asymptotic state, aligning well with observational data \cite{Iniguez:2018,Karsai:2016}. Our model extends the existing theoretical framework for social contagion modeling, which may offer valuable insights into understanding complex social contagion phenomena in reality. 

\section*{Acknowledgement}
This work was supported in part by the Key R\&D Program of Zhejiang under Grants 2022C01018 and 2024C01025, by the National Natural Science Foundation of China under Grants 62103374 and U21B2001.



\pagebreak
\clearpage
\onecolumngrid

\begin{center}
  \textbf{\large Supplementary information for social contagion under hybrid interactions}\\[.6cm]
  Xincheng Shu,$^{1,2}$ Man Yang,$^{1,2}$ Zhongyuan Ruan,$^{1,2,*}$ and Qi Xuan$^{1,2}$\\[.1cm]
  {\itshape ${}^1$Institute of Cyberspace Security, Zhejiang University of Technology,  Hangzhou 310023, China\\
  ${}^2$Binjiang Institute of Artificial Intelligence, ZJUT, Hangzhou 310056, China\\}
  ${}^*$Electronic address:  \texttt{zyruan@zjut.edu.cn}\\
\end{center}
\setcounter{equation}{0}
\setcounter{figure}{0}
\setcounter{table}{0}
\setcounter{page}{1}
\makeatletter
\renewcommand{\theequation}{S\arabic{equation}}
\renewcommand{\thefigure}{S\arabic{figure}}
\renewcommand{\bibnumfmt}[1]{[S#1]}
\renewcommand{\citenumfont}[1]{S#1}

\section{Theoretical analysis}

For each node in the network, we introduce a vector $\mathbf{k} = (k, c)$ to describe its properties, where $k = 0, 1, 2, \ldots, k_M$ is the degree of the node, and $c = 0, 1$ denotes its type, characterizing the node beyond its degree. In our model, $c = 0$ represents nodes that adopt with threshold $\phi$ (following the threshold mechanism), while $c = 1$ represents nodes that prefer to imitate others' decisions. Under the mean-field assumption, nodes with the same values of $\mathbf{k}$ are considered equivalent. We let $p_{\mathbf{k}}$ be the probability that a randomly selected node has property vector $\mathbf{k}$. Given that the imitators (a fraction of $r$) are selected randomly in the network, we have $p_{(k,0)} = (1-r)p_k$ and $p_{(k,1)} = rp_k$, where $p_k$ is the degree distribution of the underlying network.

In the updating dynamics, a susceptible node of class $\mathbf{k}=(k,0)$ adopts with probability $F_{(k,0),m}dt$ during a small time interval $dt$, where $F_{(k,0),m}$ is the infection rate, and $m$ is the number of its adopter neighbors. The rule of threshold mechanism implies that  
\begin{eqnarray}\label{eq:infection-rate1}  F_{(k,0),m}=
\begin{cases}
0, & $if $m<k\phi$$\cr 1, &$if $m\ge k\phi$$\end{cases}.
\end{eqnarray}
For a susceptible node of class $\mathbf{k}=(k,1)$, the infection rate is $F_{(k,1),m}=\theta(\beta)m/k$. We here consider the case $\theta(\beta) \to 1$.

Let $s_{\mathbf{k},m}(t)$ be the fraction of $\mathbf{k}$-class nodes that are susceptible at time $t$ and have $m$ adopting neighbors. Correspondingly, the fraction of nodes with property vector $\mathbf{k}$ that are adopters at time $t$ is $\rho_{\mathbf{k}}=1-\sum_{m=0}^k s_{\mathbf{k},m}(t)$. The total fraction of adopters in the network at time $t$ thus is given by $\rho(t)=\sum_{\mathbf{k}} p_{\mathbf{k}}\rho_{\mathbf{k}}$, where the sum over vector $\mathbf{k}$ means $\sum_{\mathbf{k}}=\sum_k\sum_c$. Following the AMEs developed by Gleeson \cite{gleeson:2013}, the rate equation for $s_{\mathbf{k},m}(t)$ can be written as 
\begin{eqnarray}\label{eq:rate-eq}  
\frac{ds_{\mathbf{k},m}}{dt}=-F_{\mathbf{k},m}s_{\mathbf{k},m}-\beta_s(k-m)s_{\mathbf{k},m}+\beta_s(k-m+1)s_{\mathbf{k},m-1},
\end{eqnarray}
where $\beta_s(t)$ is the rate at which S-S edges transform to S-I edges, $m=0,1,...,k$, and $s_{\mathbf{k},-1}\equiv 0$.  $\beta_s$ can be calculated as the ratio between the number of edges that switch from S-S to S-I during time interval $dt$ and the number of S-S edges in the network at time $t$, 
\begin{eqnarray}\label{eq:beta_s}  
\beta_s(t)=\frac{\sum_{\mathbf{k}} p_{\mathbf{k}} \sum_m (k-m)F_{\mathbf{k},m}s_{\mathbf{k},m}}{\sum_{\mathbf{k}} p_{\mathbf{k}} \sum_m (k-m) s_{\mathbf{k},m}}.
\end{eqnarray}
Assuming that a fraction $\rho(0)$ of nodes is randomly selected at the beginning, the initial conditions for Eq. (\ref{eq:rate-eq}) are $s_{\mathbf{k},m}(0)=[1-\rho(0)]B_{k,m}[\rho(0)]$, where $B_{k,m}[\rho(0)]$ is the binomial factor
\begin{eqnarray}\label{eq:binomial}  
B_{k,m}(\rho)=\binom{k}{m} \rho^m (1-\rho)^{k-m}.
\end{eqnarray}
If the initial seed size is infinitesimally small, i.e., $\rho(0) \to 0$, we have $s_{\mathbf{k},m}(0)=B_{k,m}(0)$.

For a general network with the largest degree denoted as $k_{M}$, the number of differential equations in the system (\ref{eq:rate-eq}) is $(k_M+1)(k_M+2)$, which grows quadratically with the maximum degree. Nevertheless, the AMEs for our model can be reduced to a low-dimension system by the method used by Gleeson.

To proceed, we consider three system-wide quantities: $\rho_0(t)=1-\sum_k p_k\sum_m s_{(k,0),m}(t)$, the fraction of type-0 ($c=0$) nodes that are adopters; $\rho_1(t)=1-\sum_k p_k\sum_m s_{(k,1),m}(t)$, the fraction of type-1 ($c=1$) nodes that are adopters; and $v(t)=\sum_\mathbf{k} p_{\mathbf{k}} \sum_m ms_{\mathbf{k},m}(t)/\sum_m ks_{\mathbf{k},m}(t)$, the probability that a randomly selected neighbor of a susceptible node is in adopter state. 

Note that the probability that a randomly selected node is an adopter at time $t$ can be expressed as 
\begin{eqnarray}\label{eq:rho}  
\rho(t)&=&1-\sum_{\mathbf{k}} p_{\mathbf{k}}\sum_m s_{\mathbf{k},m}(t) \nonumber\\
&=& 1- (1-r)\sum_k p_k\sum_m s_{(k,0),m}-r\sum_kp_k \sum_m s_{(k,1),m} \nonumber\\
&=& (1-r)\rho_0+r\rho_1.
\end{eqnarray}

We then propose an ansatz solution for Eq. (\ref{eq:rate-eq}) in terms of the following forms, 
\begin{eqnarray}  
s_{(k,0),m}(t)&=&B_{k,m}[v(t)] ~~~~ \text{for}~m<k\phi, \label{eq:ansatz}\\
s_{(k,1),m}(t)&=&[1-\rho_1(t)]B_{k,m}[v(t)], \label{eq:ansatz1}
\end{eqnarray}
where $B_{k,m}(v)$ follows Eq. (\ref{eq:binomial}). It is worth noticing that in Eq. (\ref{eq:ansatz1}), we have made a rather crude assumption that the probability of a randomly chosen node in class $\mathbf{k}=(k,1)$ being susceptible is $1-\rho_1(t)$. This assumption is applicable when considering a homogeneous underlying network. 

Next, we insert the ansatz (\ref{eq:ansatz}) into the AME system (\ref{eq:rate-eq}) for $m<k\phi$, hoping to get a set of differential equations for quantities $\rho_0$, $\rho_1$ and $v$. Taking the time derivative of Eq. (\ref{eq:ansatz}), we get 
\begin{eqnarray}\label{eq:der1}  
\dot{s}_{(k,0),m}=(\frac{m}{v}-\frac{k-m}{1-v})\dot{v} s_{(k,0),m}.
\end{eqnarray}
Using the binomial identity 
\begin{eqnarray}\label{eq:identity}  
B_{k,m-1}=\frac{1-v}{v}\frac{m}{k-m+1}B_{k,m}(v),
\end{eqnarray}
the ansatz (\ref{eq:ansatz}), and the threshold rule (\ref{eq:infection-rate1}) for $m<k\phi$, we obtain from the right hand side of Eq. (\ref{eq:rate-eq}) that
\begin{eqnarray}\label{eq:righthand}  
-F_{(k,0),m}s_{(k,0),m}-\beta_s(k-m)s_{(k,0),m}+\beta_s(k-m+1)s_{(k,0),m-1}=\beta_s(m-k+\frac{1-v}{v}m)s_{(k,0),m}.
\end{eqnarray}
Comparing Eqs.(\ref{eq:der1}) and (\ref{eq:righthand}), we obtain the condition on $v$ so that the ansatz (\ref{eq:ansatz}) is a solution of Eq. (\ref{eq:rate-eq}),
\begin{eqnarray}\label{eq:dotv}  
\dot{v}=\beta_s(1-v).
\end{eqnarray}
The initial condition of Eq. (\ref{eq:dotv}) is $v(0)=\rho(0)=0$ (by comparing the initial condition of $s_{\mathbf{k},m}$ and the ansatz for $t=0$). Furthermore, we assume a function $g(v, \rho_0,\rho_1)$ which satisfies
\begin{eqnarray}\label{eq:g_v}  
\dot{v}=g(v,\rho_0,\rho_1)-v.
\end{eqnarray}
By comparing Eqs. (\ref{eq:dotv}) and (\ref{eq:g}), we get 
\begin{eqnarray}\label{eq:beta_s2}  
\beta_s=\frac{g(v,\rho_0,\rho_1)-v}{1-v}.
\end{eqnarray}

The next step is to determine the exact expression for the function $g(v,\rho_0,\rho_1)$. We need to apply the following general result derived by Gleeson in \cite{gleeson:2013}, which holds for any $F_{\mathbf{k},m}$,
\begin{eqnarray}\label{eq:general}  
\sum_{\mathbf{k}} p_{\mathbf{k}} \sum_m(k-m)s_{{\mathbf{k}},m}=z(1-v)^2,
\end{eqnarray}
with $z=\sum_k kp_k$, being the average degree of the network. Therefore, Eq. (\ref{eq:beta_s}) can be rewritten as
\begin{eqnarray}\label{eq:beta_s3}  
\beta_s&=&\frac{1}{z(1-v)^2} \sum_{\mathbf{k}} p_{\mathbf{k}}\sum_m (k-m)F_{\mathbf{k},m} s_{\mathbf{k},m} \nonumber\\
&=& \frac{1}{z(1-v)^2} \left[ \sum_k (1-r)p_k\sum_{m\ge k\phi}(k-m)s_{(k,0),m}+\sum_k rp_k\sum_m(k-m)\frac{m}{k}s_{(k,1),m}  \right] \nonumber\\
&=& \frac{1}{z(1-v)^2} \left[ \sum_{\mathbf{k}} p_{\mathbf{k}}\sum_m(k-m)s_{\mathbf{k},m}-(1-r)\sum_kp_k\sum_{m<k\phi}(k-m)s_{(k,0),m}-r\sum_kp_k\sum_{m}\frac{(k-m)^2}{k}s_{(k,1),m} \right] \nonumber\\
&=&\frac{1}{z(1-v)^2} \left[ z(1-v)^2-(1-r)\sum_kp_k\sum_{m<k\phi}(k-m)B_{k,m}(v)-r(1-\rho_1)\sum_kp_k\sum_{m}\frac{(k-m)^2}{k}B_{k,m}(v) \right],
\end{eqnarray}
where we have used the ansatz (\ref{eq:ansatz}) and (\ref{eq:ansatz1}), as well as Eq. (\ref{eq:general}) in the last line. Using the identity $(k-m)B_{k,m}(v)=k(1-v)B_{k-1,m}(v)$  [and $(k-m-1)B_{k-1,m}(v)=(k-1)(1-v)B_{k-2,m}(v)$], $\beta_s$ can be further rewritten as
\begin{eqnarray}\label{eq:beta_s4}  
\beta_s&=&\frac{1}{1-v} \left[1-v-(1-r)\sum_k \frac{kp_k}{z}\sum_{m<k\phi}B_{k-1,m}(v)-r(1-\rho_1)(1-v+\frac{v}{z}) \right] \nonumber\\
&=&\frac{1}{1-v} \left[ \left((1-r)\sum_k \frac{kp_k}{z}\sum_{m\ge k\phi}B_{k-1,m}(v)+r[1-(1-\rho_1)(1-v+\frac{v}{z})]\right)-v \right].
\end{eqnarray}

Comparing Eqs. (\ref{eq:beta_s4}) and (\ref{eq:beta_s2}), we  get the expression for $g(v,\rho_0,\rho_1)$, 
\begin{eqnarray}\label{eq:g}  
g(v,\rho_0,\rho_1)=(1-r)\sum_k \frac{kp_k}{z}\sum_{m\ge k\phi}B_{k-1,m}(v)+r[1-(1-\rho_1)(1-v+\frac{v}{z})],
\end{eqnarray}
which is independent of $\rho_0$. Furthermore, we can also derive the evolution equation for $\rho_0$ and $\rho_1$. From the definition of $\rho_c$ ($c=0,1$) and Eq.(\ref{eq:rate-eq}), we obtain
\begin{eqnarray}\label{eq:rho_c}  
\dot{\rho}_c&=&-\sum_k p_k\sum_m \dot{s}_{(k,c),m} \nonumber\\
&=&\sum_k p_k\sum_m F_{(k,c),m}s_{(k,c),m}+\beta_s \sum_k p_k\sum_m [(k-m)s_{(k,c),m}-(k-m+1)s_{(k,c),m-1}].
\end{eqnarray}
Note that the second term in the right-hand side telescopes to $0$. Thus we have 
\begin{eqnarray}\label{eq:rho0}  
\dot{\rho}_0&=&\sum_k p_k\sum_m F_{(k,0),m}s_{(k,0),m}=\sum_k p_k\sum_{m\ge k\phi} s_{(k,0),m}\nonumber\\
&=&\sum_k p_k\sum_{m} s_{(k,0),m}-\sum_k p_k \sum_{m<k\phi} s_{(k,0),m} \nonumber\\
&=&1-\rho_0-\sum_k p_k \sum_{m<k\phi} B_{k,m}(v)\nonumber\\
&=&1-\rho_0-\sum_k p_k [1-\sum_{m\ge k\phi}B_{k,m}(v)]\nonumber\\
&=&\left(\sum_k p_k\sum_{m\ge k\phi}B_{k,m}(v)\right)-\rho_0.
\end{eqnarray}
Define 
\begin{eqnarray}\label{eq:h_v}  
h(v)=\sum_k p_k\sum_{m\ge k\phi}B_{k,m}(v),
\end{eqnarray}
Eq.(\ref{eq:rho0}) can be expressed as $\dot{\rho_0}=h(v)-\rho_0$. Similarly, we have

\begin{eqnarray}\label{eq:rho1}  
\dot{\rho}_1&=&\sum_k p_k\sum_m F_{(k,1),m}s_{(k,1),m}=\sum_k p_k\sum_m \frac{m}{k}(1-\rho_1) B_{k,m}(v)\nonumber\\
&=&(1-\rho_1)\sum_k p_k \sum_m v B_{k-1,m-1}(v)\nonumber\\
&=&(1-\rho_1)v.
\end{eqnarray}

To summarize, Eq. (\ref{eq:rate-eq}) can be reduced to three ordinary differential equations
\begin{eqnarray}\label{eq:reduced}  
\dot{\rho}_0&=&h(v)-\rho_0, \\
\dot{\rho}_1&=&(1-\rho_1)v, \\
\dot{v}&=&g(v,\rho_1)-v, 
\end{eqnarray}
with the functions $g(v)$ and $h(v)$ given explicitly by Eqs. (\ref{eq:g}) and (\ref{eq:h_v}). The above equations can be solved numerically to obtain $\rho_0(t)$, $\rho_1(t)$ and $v(t)$. Finally, the fraction of adopters in the network $\rho(t)$ can be given by Eq.(\ref{eq:rho}).


\end{document}